\documentclass[preprintnumbers,amsmath,11pt,amssymb,floatfix,superscriptaddress,nofootinbib]{revtex4}
\usepackage{graphicx}
\usepackage{epsfig}
\usepackage{bm}
\usepackage{amsfonts}

\input epsf

\begin{document}

\title{New HDE models with higher derivatives of the Hubble parameter $H$}
\author{Antonio Pasqua}
\email{toto.pasqua@gmail.com}

\date{\today}
\newpage

%\begin{document}

\begin{abstract}
In this work, we investigate two Dark Energy (DE) models characterized by higher-order derivatives of the Hubble parameter $H$, which generalize previously proposed DE scenarios. Assuming a power-law form of the scale factor $a(t)$ given by $a(t)=b_0t^n$, we derive analytical expressions for the DE energy density, pressure, the Equation of State (EoS) parameter, the deceleration parameter and the evolutionary form of the fractional DE density. Both non-interacting and interacting dark sector frameworks are examined, with the interaction modeled through a coupling term proportional to the Dark Matter (DM) energy density. 

For specific parameter sets corresponding to power-law indices $n=2$, $n=3$, and $n=4$, we compute the present age of the Universe. The values obtained slightly deviate from the observationally inferred age of $\approx 13.8$ Gyr; moreover, a systematic trend is identified, with larger $n$ leading to higher ages. Furthermore, interacting scenarios consistently predict larger ages compared to their non-interacting counterparts. These results highlight the phenomenological viability and limitations of higher-derivative DE models in describing the cosmic evolution.
\end{abstract}

\maketitle
\tableofcontents
\section{Introduction}
Observations from the Wilkinson Microwave Anisotropy Probe (WMAP) \cite{cmb1,cmb2}, the Supernova Cosmology Project \cite{sn2,sn4}, the Sloan Digital Sky Survey (SDSS) \cite{sds1,sds3,sds4}, the Planck mission \cite{planck}, and X-ray studies \cite{xray} consistently indicate that the Universe is currently undergoing accelerated expansion. To explain this phenomenon, a hypothetical component called Dark Energy (DE), with a negative pressure, has been introduced. The Cosmological Constant $\Lambda_{CC}$ is the simplest candidate, yet it faces conceptual challenges such as the cosmological constant and coincidence problems, motivating the development of alternative dynamical models \cite{cosm3,cosm4,cosm5}.

From a theoretical perspective, $\Lambda_{CC}$ can be incorporated as a constant term in Einstein's field equations. Quantum Field Theory calculations using Planck- or electroweak-scale cut-offs predict vacuum energy densities vastly larger than observed, by factors of $10^{123}$ and $10^{55}$, respectively. The lack of a natural symmetry to suppress $\Lambda_{CC}$ gives rise to the cosmological constant problem, while the coincidence problem questions why matter and DE densities are comparable today \cite{cosm3,cosm4,cosm5,rev2}.  

Within the standard cosmological model, DE accounts for approximately two-thirds of the current energy density $\rho_{\text{tot}}$ \cite{twothirds}, with the remaining fraction composed mainly of Dark Matter (DM) and baryons. Despite precise measurements, the microscopic origin of DE remains elusive.

Dynamical DE models, where the equation of state parameter $\omega_D$ evolves over time, provide a flexible framework consistent with observational data. Examples include scalar field scenarios such as quintessence \cite{quint1,quint4,quint5}, k-essence \cite{kess3,kess4,kess5}, tachyon \cite{tac2,tac3,tac4}, phantom \cite{pha2,pha5,pha6}, dilaton \cite{dil1,dil2}, and quintom models \cite{qui3,qui8,qui10,qui12}. Interacting DE models, including those based on the Chaplygin gas \cite{cgas1,cgas2,cgas3}, the Agegraphic Dark Energy (ADE) and the New ADE (NADE) \cite{ade1,ade2}, have also been explored.  

Holographic approaches provide a complementary perspective, based on the principle that the entropy of a system scales with its boundary area rather than volume \cite{holo1,holo2,holo5}. Holographic Dark Energy (HDE), first proposed by Li \cite{li}, postulates a DE density
\begin{equation}
\rho_D= 3\alpha M_p^2 L^{-2},
\end{equation}
where $\alpha$ is a dimensionless constant and $M_p = (8\pi G_N)^{-1/2}$ is the reduced Planck mass. Cohen et al.~\cite{coh1} initially suggested that the vacuum energy should be bounded to prevent black hole formation, but the naive choice $\rho_\Lambda \propto H^2$ fails to drive acceleration. Using the future event horizon as the infrared cutoff, HDE models successfully reproduce late-time cosmic acceleration \cite{li}. Extensions include the Holographic Ricci Dark Energy, with $L \propto R^{-1/2}$ \cite{gaoprimo}, and the Granda–Oliveros formulation, where $\rho_\Lambda$ depends on both $H$ and $\dot{H}$ \cite{go1,go4,go5}. These models have been extensively confronted with supernova, CMB, and BAO observations \cite{cons1,cons2,cons3,cons4,cons5,cons6,cons7,cons8,cons9} and are reviewed in \cite{hde1,hde2,hde7,hde10,hde12,YYS,hde13,hde17,hde18,hde19,
hde22,hde23,hde24,hde26,hde28,hde30,hde32,hde33,hde34,hde35,saridakis11,saridakis22}.

We now want to study two different Dark Energy energy density models which involve higher time derivatives of the Hubble parameter $H$. \\
The first one is given by:
\begin{eqnarray}
\rho_D = 3 \left[ \alpha \left(\frac{\dddot{H}}{H^2}\right)  +\beta \left(\frac{\ddot{H}}{H}\right)  + \gamma \dot{H}+\delta H^2\right] ,\label{cz1}
\end{eqnarray}
where the quantities $\alpha$, $\beta$, $\gamma$ and $\delta$ are dimensionless parameters.  For mathematical simplicity, we set the reduced Planck mass to unity, i.e. $M_p = 1$. We note that the inverse of the Hubble parameter squared $H^{-2}$ and the inverse of the Hubble parameter $H^{-1}$ are included in the first and second terms to guarantee consistency of physical dimensions across all terms.

The cosmological behavior and main features of this DE model depend crucially on the four parameters of the model. The energy density given in Eq.~(\ref{cz1}) can be regarded as a generalization of several previously proposed DE models. For instance, by setting $\alpha = 0$, we recover the energy density introduced in Chen \& Jing \cite{modelhigher} and studied in other subsequent works. Instead, in the limiting case of $\alpha = \beta=0$, the energy density given in Eq. (\ref{cz1}) reduces to the energy density of DE with the Granda-Oliveros cut-off \cite{gohnde}. Furthermore, for the particular choice $\alpha=\beta=0$, $\gamma=1$, and $\delta=2$, the model reproduces the dark energy (DE) density with infrared (IR) cut-off determined by the average radius associated with the Ricci scalar, valid in a spatially flat Universe ($k=0$). Since the present model introduces an additional free parameter, it provides a more general framework than the Ricci Dark Energy (RDE) scenario. Similar DE models have been investigated in detail in \cite{altri3,altri1,altri2}.\\
The general expression of the second model we consider is given by:
\begin{eqnarray}
\rho_D = 3 \left[ \alpha \left(\frac{\dddot{H}}{H^2}\right) - \zeta\left(  \frac{\ddot{H}\dot{H}}{H^3}  \right)  +\beta \left(\frac{\ddot{H}}{H}\right)  + \gamma \dot{H}+\delta H^2\right] ,\label{cz2}
\end{eqnarray}
where $\alpha$, $\beta$, $\gamma$, $\delta$ and $\zeta$ are five constant parameters. In the limiting case of $\zeta=0$, we recover the first model introduced.\\
Moreover, in the limiting case of $\alpha=\zeta=0$, we recover the model studied in Chen \& Jing \cite{modelhigher}, while for $\alpha=\beta=\zeta=0$ we obtain the HDE energy density model with Granda-Oliveros cut-off. Furthermore, for  $\alpha=\beta=\zeta=0$, $\gamma=1$ and $\delta=2$ we obtain the HDE model with cut-off given by the average radius of the Ricci scalar for a spatially flat Universe. \\
In the following sections, we derive key cosmological quantities for a power-law form of the scale factor in these two models, including the dark energy density, dark energy pressure, the equation of state parameter, the deceleration parameter, and the evolution of the dark energy fractional density. Additionally, we compute the age of the Universe for selected values of the relevant parameters.\\
The paper is structured as follow. In Section 2, we study the first model introduced.  In Section 3, we study the second model we consider. In Section 4, we evaluate the present age of the Universe for both models, exploring different sets of model parameters. 
Finally, in Section 5, we write the Conclusions of this work.

\section{Holographic Dark Energy Model in a Non-Flat Universe}
In this Section, we describe the main features of the first Dark Energy (DE) model under investigation in this paper and derive some fundamental cosmological quantities. \\
The geometry of a Universe assumed to be homogeneous and isotropic is represented by the Friedmann–Lemaitre–Robertson–Walker (FLRW) metric, expressed as:
\begin{eqnarray}
    ds^2 &=&-dt^2 + a^2(t)\left[\frac{dr^2}{1 - kr^2} + r^2d\Omega^2\right]\nonumber \\
    &=&-dt^2 + a^2(t)\left[\frac{dr^2}{1 - kr^2} + r^2 (d\theta^2 + \sin^2\theta\, d\varphi^2)\right], \label{6}
\end{eqnarray}
where $t$ represents the cosmic time, $a(t)$ indicates the scale factor (which describes the expansion of the Universe), the coordinate $r$ indicates the comoving radial coordinate, and $\theta$ and $\varphi$ are the two usual angular coordinates in spherical symmetry, which can assume values $0 \leq \theta \leq \pi$ and $0 \leq \varphi < 2\pi$. The parameter $k$ denotes the spatial curvature and may take the values 
-1, 0 and +1 which correspond to open, flat, and closed Universes, respectively.\\
The evolution of a homogeneous and isotropic Universe within the framework of General Relativity is determined by the Friedmann equations, which, in the presence of both DE and DM, can be written as:
\begin{eqnarray}
    H^2  &=& \frac{1}{3M^2_p}\left( \rho_D + \rho_m \right)-\frac{k}{a^2}, \label{7} \\
    \dot{H} + 2H^2 &=& \left(\frac{8\pi G}{6}\right)\, p_D-\frac{k}{a^2}, \label{7fri2}
\end{eqnarray}
where $H = \dot{a}/a$ represents the Hubble parameter, $\rho_D$ is the energy density of DE while $p_D$ denote the pressure of DE. Moreover, we have that $\rho_m$ represents the energy density of DN. \\
We define the expressions of the fractional energy densities  DE, matter and curvature as follows:
\begin{eqnarray}
    \Omega_D &=& \frac{\rho_D}{\rho_{\text{cr}}} = \frac{\rho_D}{3M^2_p H^2}, \label{10} \\
    \Omega_m &=& \frac{\rho_m}{\rho_{\text{cr}}} = \frac{\rho_m}{3M^2_p H^2}, \label{8} \\
    \Omega_k &=& \frac{k}{a^2 H^2}. \label{10k}
\end{eqnarray}
The quantity $\rho_{\text{cr}}$ indicates the critical energy density required for a spatially flat Universe, and it can be written as:
\begin{eqnarray}
    \rho_{\text{cr}} = 3M^2_p H^2.
\end{eqnarray}

Using the expression for $\Omega_D$ and $\Omega_m$ given in Eqs.~(\ref{10}) and (\ref{8}), the Friedmann equation obtained in Eq.~(\ref{7}) can be written in the following form:
\begin{eqnarray}
    \Omega_D + \Omega_m + \Omega_k = 1. \label{11}
\end{eqnarray}
To guarantee the fulfillment of the Bianchi identity, or equivalently the local conservation of energy-momentum, the total energy density $\rho_{\text{tot}} = \rho_D + \rho_m$ must satisfy the continuity equation:
\begin{eqnarray}
    \dot{\rho}_{\text{tot}} + 3H\left( \rho_{\text{tot}} + p_{\text{tot}} \right) = 0, \label{12old}
\end{eqnarray}
where the quantities $\rho_{\text{tot}}$ and $p_{\text{tot}}$ are the total energy density and the total pressure of the cosmic fluid, respectively, and they are given by the following relations:
\begin{eqnarray}
    \rho_{\text{tot}} &=& \rho_m + \rho_D, \\
    p_{\text{tot}} &=& p_D,
\end{eqnarray}
since we assume that DM is pressureless, i.e. we consider $p_m=0$.\\
The expression of the continuity equation we obtained in Eq.~(\ref{12old}) can be also expressed as a function  of the total EoS parameter $\omega_{\text{tot}} = p_{\text{tot}} / \rho_{\text{tot}}$ as follows:
\begin{eqnarray}
    \dot{\rho}_{\text{tot}} + 3H\left(1 + \omega_{\text{tot}} \right) \rho_{\text{tot}} = 0. \label{12}
\end{eqnarray}

Since the energy densities of DM and DE are assumed to be conserved separately, we have that Eq.~(\ref{12}) can be decomposed into two independent continuity equations. In the non-interacting case, these read:
\begin{eqnarray}
    \dot{\rho}_D + 3H\left( 1 + \omega_D \right)\rho_D &=& 0, \label{12deold} \\
    \dot{\rho}_m + 3H\rho_m &=& 0. \label{12dm}
\end{eqnarray}
Considering the general definition for the EoS parameter $\omega_D$ for DE given by:
\begin{eqnarray}
    \omega_D = \frac{p_D}{\rho_D}, \label{mammud}
\end{eqnarray}
it is possible to rewrite Eq.~(\ref{12deold}) in the following form:
\begin{eqnarray}
    \dot{\rho}_D + 3H\left( p_D + \rho_D \right) = 0. \label{12de}
\end{eqnarray}

The results obtained i Eqs.~(\ref{12deold}), (\ref{12dm}) and (\ref{12de}) can be also rewritten as functions of the variable $x$ defined as $x = \ln a$, using the following relation for the derivative: $\frac{d}{dt} = H \frac{d}{dx}$. Denoting the derivatives with respect to the variable $x$ by a prime, we obtain the following results:
\begin{eqnarray}
    \rho'_D + 3\left( 1 + \omega_D \right)\rho_D &=& 0, \label{12deoldprime} \\
    \rho'_D + 3\left( p_D + \rho_D \right) &=& 0, \label{12deprime} \\
    \rho'_m + 3\rho_m &=& 0. \label{12dmprime}
\end{eqnarray}

At present, dark energy (DE) represents approximately two-thirds of the Universe’s total energy density, whereas its effect was negligible in the very early epochs. This observation suggests that DE is not static but evolves with cosmic expansion. In this framework, it is natural to examine models where the density of DE depends on the Hubble rate $H$ and its time derivatives, which encode the dynamics of cosmic expansion.

\section{First Model}
We now study the first model of energy density considered in this paper.\\
In this work, we consider a power-law form of the scale factor $a(t)$ given by the following relation:
\begin{equation}
a(t) = b_0 t^n, \label{scale}
\end{equation}
where \(b_0\) and \(n\) represent two positive constants. We must underline we consider $n>0$. We will make some considerations about the value of $b_0$ later in the paper. \\
We now want to calculate some quantities using the expression of the scale factor given in Eq. (\ref{scale}).\\
Using Eq. (\ref{scale}) in the expression of $\rho_D$ given in Eq. (\ref{cz1}), we obtain the following expression for $\rho_D$ as function of the time:
\begin{eqnarray}
\rho_{D,1}(t) 
= \frac{3}{t^2} \left( -\frac{6\alpha}{n} + 2\beta - \gamma n + \delta n^2 \right).\label{cz3}
\end{eqnarray}
We now want to calculate the expression of the EoS paramater for DE.\\
From the continuity equation given in Eq. (\ref{12deold}), we find the general expression:
\begin{eqnarray}
\omega_{D,1} = -1 - \frac{\dot{\rho}_{D,1}}{3H\rho_{D,1}}  .  
\end{eqnarray}

From Eq. (\ref{cz3}), we obtain:
\begin{eqnarray}
\dot{\rho}_{D,1}(t) 
= -\frac{6}{t^3} \left( -\frac{6\alpha}{n} + 2\beta - \gamma n + \delta n^2 \right).
\end{eqnarray}
Moreover, we obtain that the Hubble parameter as function of the time is given by:
\begin{eqnarray}
    H(t) = \frac{\dot{a}}{a}=\frac{n}{t}.
\end{eqnarray}
Therefore, we obtain the following relation for $\omega_D$:
\begin{eqnarray}
    \omega_{D,1} = -1+\frac{2}{3n} .\label{eos}
\end{eqnarray}
This expression implies that $\omega_{D,1}$ is always greater than $-1$, 
indicating a quintessence-like behavior. In particular, for small values of $n$,  the EoS parameter deviates significantly from the cosmological constant limit,  reflecting a less dominant dark energy component. 
For $n=1$, the equation of state (EoS) parameter of dark energy is 
\begin{eqnarray}
\omega_{D,1} = -\frac{1}{3}.
\end{eqnarray}
This value corresponds to the \textit{critical limit} between decelerated and accelerated expansion of the Universe:
\begin{itemize}
    \item If $\omega_{D,1} > -1/3$ (i.e. $< n <1$), the expansion is decelerating.
    \item If $\omega_{D,1} < -1/3$, (i.e. ($n>1$)) the expansion is accelerating.
\end{itemize}
Thus, for $n=1$, the model predicts a marginal expansion at the boundary between acceleration and deceleration.

\begin{figure}[htbp]
    \centering
    \includegraphics[width=0.7\textwidth]{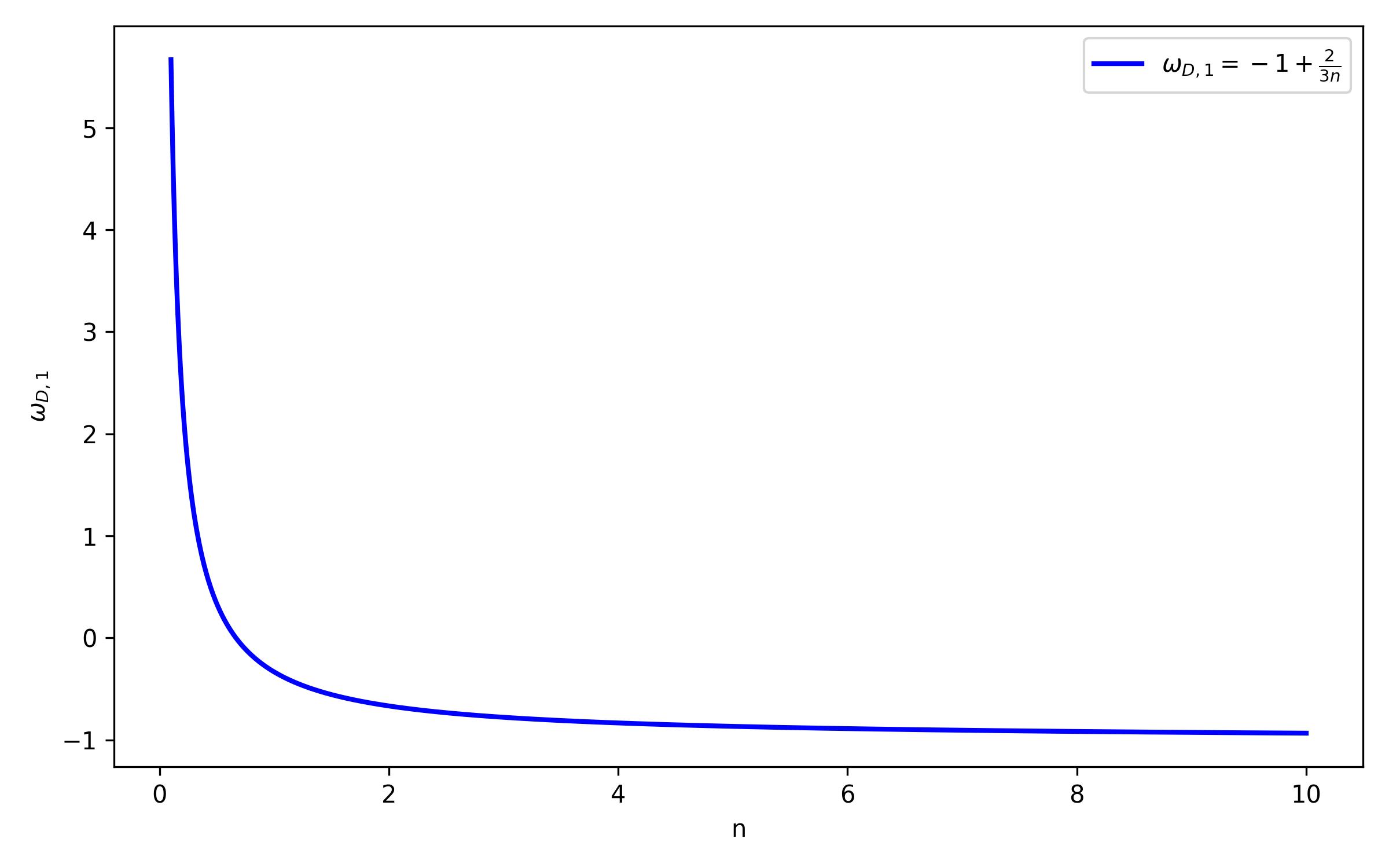} 
    \caption{Plot of the expression of $\omega_{D,1}$ for $n \in \left[0.1-10\right] $.}
    \label{1}
\end{figure}

As $n$ increases, $\omega_{D,1}$ approaches $-1$, effectively mimicking a cosmological constant. 
Therefore, the parameter $n$ controls the deviation of dark energy from a pure  $\Lambda$-like behavior, with higher values of $n$ corresponding to an accelerated  expansion that closely resembles a $\Lambda$CDM scenario. 
Overall, the model predicts a non-phantom, accelerating universe driven by  quintessence-like dark energy.\\
In Fig. (\ref{1}) we plot the behavior of $\omega_{D,1}$ obtained in Eq. (\ref{eos})  in the range of values of $n \in \left[0.1-10\right] $.
We now derive the expression of the pressure of DE $p_{D,1}$.\\
From the continuity equation for DE, we obtain the following general expression for $ p_{D,1}$:
\begin{eqnarray}
    p_{D,1}= -\rho_{D,1} - \frac{\dot{\rho}_{D,1}}{3H}.
\end{eqnarray}
Using the expressions of $\rho_{D,1}$, $\dot{\rho}_{D,1}$ and $H$ we derived before, we can write: 
\begin{eqnarray}
    p_{D,1}(t) &=& \left( \frac{2}{3n} -1 \right)\rho_{D,1}(t)\nonumber \\
    &=& \left( \frac{2}{n} -3 \right)\left( -\frac{6\alpha}{n} + 2\beta - \gamma n + \delta n^2 \right)t^{-2}.
\end{eqnarray}

We now want to obtain the expression of the Hubble parameter squared as function of the time.
For the energy density of DM, from the continuity equation obtain in Eq. (\ref{12dm}), we obtain:
\begin{eqnarray}
    \rho_m= \rho_{m0}a^{-3}.
\end{eqnarray}
Using the expression of $a(t)$ we have chosen in this paper, we can write:
\begin{eqnarray}
    \rho_m&=& \rho_{m0}b_0^{-3}t^{-3n}\nonumber \\
    &=&\frac{\rho_{m0}}{b_0^{3}t^{3n}},
\end{eqnarray}
which leads to
\begin{eqnarray}
    \rho_m&=& 3H_0^2\Omega_{m0}b_0^{-3}t^{-3n}\nonumber \\
    &=&\frac{3H_0^2\Omega_{m0}}{b_0^{3}t^{3n}},
\end{eqnarray}
where we used the fact that $\rho_{m0}=3H_0^2\Omega_{m0}$.\\
Generally, for the curvature energy density, we have:
\begin{eqnarray}
    \rho_k= ka^{-2}.
\end{eqnarray}
Using the expression of the scale factor we consider in this paper, we can write:
\begin{eqnarray}
    \rho_k&=& k\cdot b_0^{-2}t^{-2n}\nonumber \\
    &=&\frac{k}{b_0^{2}t^{2n}},
\end{eqnarray}
which leads to:
\begin{eqnarray}
    \rho_k&=& H_0^2\Omega_{k0}\cdot b_0^{-2}t^{-2n}\nonumber \\
    &=&\frac{H_0^2\Omega_{k0}}{b_0^{2}t^{2n}},
\end{eqnarray}
where we used the fact that $k= H_0^2\Omega_{k0}$.\\
We then obtain the following expression for $H^2_1(t)$:
\begin{eqnarray}
    H_1^2(t)= H_0^2\left\{  \frac{\Omega_{m0}}{b_0^{3}t^{3n}}+\frac{\Omega_{k0}}{b_0^{2}t^{2n}}+\frac{1}{H_0^2t^2} \left( -\frac{6\alpha}{n} + 2\beta - \gamma n + \delta n^2 \right)\right\}.
\end{eqnarray}

We now want to write the equations we derived as functions of the redshift $z$.\\
The energy density of DE can be written as:
\begin{eqnarray}
\rho_{D,1}(z) \;=\; 3\,
\left(-\frac{6\alpha}{n}+2\beta-\gamma n+\delta n^2\right)b_0^{2\over n}\,(1+z)^{2\over n}.
\end{eqnarray}

The pressure of DE $p_{D,1}$ is given by:
\begin{eqnarray}
p_{D,1}(z) &=& \left(\frac{2}{3n}-1\right) \rho_D(z) \nonumber\\
&=& \left(\frac{2}{n}-3\right) 
 \left(-\frac{6\alpha}{n} + 2\beta - \gamma n + \delta n^2 \right) b_0^{2\over n} (1+z)^{2\over n}.
\end{eqnarray}

The Hubble parameter squared as function of $z$ is given by:
\begin{eqnarray}
    H_1^2(z)= H_0^2\left\{  \Omega_{m0}(1+z)^3+\Omega_{k0}(1+z)^2+ \frac{1}{H_0^2} \left( -\frac{6\alpha}{n} + 2\beta - \gamma n + \delta n^2 \right)b_0^{2\over n}\,(1+z)^{2\over n}\right\}.\label{unical1}
\end{eqnarray}
Finally, we can obtain the deceleration parameter $q$ from the relation:
\begin{eqnarray}
    q_1(z)= -1+\left(\frac{1+z}{2}\right)\frac{1}{h_1^2(z)}\frac{dh_1^2(z)}{dz},
\end{eqnarray}
where $h_1^2=H_1^2/H_0^2$
From the result of Eq. (\ref{unical1}), we can write:
\begin{eqnarray}
    \frac{dh_1^2(z)}{dz} =  3\Omega_{m0}(1+z)^2+2\Omega_{k0}(1+z)+ \frac{2}{nH_0^2} \left( -\frac{6\alpha}{n} + 2\beta - \gamma n + \delta n^2 \right)b_0^{2\over n}\,(1+z)^{\frac{2}{n}-1}.
\end{eqnarray}
Therefore, the final expression of $q_1(z)$ is given by:
\begin{eqnarray}
    q_1(z)&=&-1+\frac{1}{2}\cdot \left[  3\Omega_{m0}(1+z)^3+2\Omega_{k0}(1+z)^2+ \frac{2}{nH_0^2} \left( -\frac{6\alpha}{n} + 2\beta - \gamma n + \delta n^2 \right)b_0^{2\over n}\,(1+z)^{\frac{2}{n}}  \right]\times \nonumber \\
    &&\left[  \Omega_{m0}(1+z)^3+\Omega_{k0}(1+z)^2+ \frac{1}{H_0^2} \left( -\frac{6\alpha}{n} + 2\beta - \gamma n + \delta n^2 \right)b_0^{2\over n}\,(1+z)^{\frac{2}{n}}  \right].
\end{eqnarray}

We can now make some considerations about $b_0$.\\
Considering the powe law form of the scale factor we defined in Eq. (\ref{scale}), we obtain that the Hubble parameter can be written as follows: 
\begin{eqnarray}
H(t) = \frac{n}{t}.
\end{eqnarray}
If we adopt the conventional normalization $a_0=1$, then the present time $t_0$ satisfies
\begin{eqnarray}
1 = a_0 = b_0 t_0^n \;\;\Rightarrow\;\; t_0 = b_0^{-1/n}.
\end{eqnarray}
Therefore,
\begin{eqnarray}
H_0 = H(t_0) = \frac{n}{t_0} = n\,b_0^{1/n}. \label{unical2}
\end{eqnarray}

From Eq. (\ref{unical2}), we obtain
\begin{eqnarray}
b_0=\left(\frac{H_0}{n}\right)^n \rightarrow b_0^{2\over n} = \left(\frac{H_0}{n}\right)^{2}.
\end{eqnarray}

Using the result we obtained for $b_0^{2\over n}$, the energy density of DE assumes the following form:
\begin{eqnarray}
\rho_{D,1}(z) &=& 3\left(-\frac{6\alpha}{n} + 2\beta - \gamma n + \delta n^2\right) \left(\frac{H_0}{n}\right)^{2}(1+z)^{2\over n}\nonumber \\
&=&3\left(-\frac{6\alpha}{n^3} + \frac{2\beta}{n^2} - \frac{\gamma}{n}+ \delta \right) H_0^{2}(1+z)^{2\over n}.
\end{eqnarray}

The Hubble parameter squared can be written as:
\begin{eqnarray}
    H_1^2(z)= H_0^2\left\{  \Omega_{m0}(1+z)^3+\Omega_{k0}(1+z)^2+ \left(-\frac{6\alpha}{n^3} + \frac{2\beta}{n^2} - \frac{\gamma}{n}+ \delta \right) \,(1+z)^{2\over n}\right\},
\end{eqnarray}
while the pressure of DE can be rewritten as:
\begin{eqnarray}
p_{D,1}(z) &=& \left( \frac{2}{3n}-1\right) \rho_D(z) \nonumber\\
&=& \left( \frac{2}{n}-3\right) 
 \left(-\frac{6\alpha}{n^3} + \frac{2\beta}{n^2} - \frac{\gamma}{n}+ \delta \right)H_0^2(1+z)^{2\over n}.
\end{eqnarray}

Finally, the deceleration parameter is given by:
\begin{eqnarray}
    q_1(z)&=&-1+\frac{1}{2}\cdot \left[  3\Omega_{m0}(1+z)^3+2\Omega_{k0}(1+z)^2+ \frac{2}{n} \left(-\frac{6\alpha}{n^3} + \frac{2\beta}{n^2} - \frac{\gamma}{n}+ \delta \right)\,(1+z)^{\frac{2}{n}}  \right]\times \nonumber \\
    &&\left[  \Omega_{m0}(1+z)^3+\Omega_{k0}(1+z)^2+  \left(-\frac{6\alpha}{n^3} + \frac{2\beta}{n^2} - \frac{\gamma}{n}+ \delta \right)\,(1+z)^{\frac{2}{n}}  \right]^{-1}.
\end{eqnarray}

The present day value of $q_1$ is given by:
\begin{eqnarray}
    q_{1,0}&=&-1+\frac{1}{2}\cdot \left[  3\Omega_{m0}+2\Omega_{k0}+ \frac{2}{n}\left(-\frac{6\alpha}{n^3} + \frac{2\beta}{n^2} - \frac{\gamma}{n}+ \delta \right)  \right]\times \nonumber \\
    &&\left[  \Omega_{m0}+\Omega_{k0}+  \left(-\frac{6\alpha}{n^3} + \frac{2\beta}{n^2} - \frac{\gamma}{n}+ \delta \right) \right]^{-1}.
\end{eqnarray}

We now want to find the evolutionary form of the fractional energy density of DE for this model.\\
Using the expression of $\rho_{D,1}$, we can write:
\begin{eqnarray}
\Omega_{D,1} &=& \frac{\rho_{D,1}}{3H^2}\nonumber\\
&=&\left[-\frac{6\alpha}{n^3} + \frac{2\beta}{n^2} - \frac{\gamma}{n}+ \delta \right],
\end{eqnarray}
i.e. $\Omega_{D,1}$ is a constant depending on the values of the parameters of the energy density and on the power law index of the scale factor we have chosen. \\
Therefore, the evolutionary form of $\Omega_{D,1}$ is $\Omega_{D,1}'=0$, i.e. we have that the fraction of DE relative to the total critical density does not change over time.\\
We have then that the present day value of $\Omega_{D,1}$ is given by:
\begin{eqnarray}
\Omega_{D,1,0}
&=&\left[-\frac{6\alpha}{n^3} + \frac{2\beta}{n^2} - \frac{\gamma}{n}+ \delta \right].
\end{eqnarray}
We can obtain some hints about the values of the parameters of the model using the value of $\Omega_{D,1,0}$. \\
We now want that, at present time,  $\Omega_{D,1,0}\approx 0.685$, therefore we should have:
\begin{eqnarray}
-\frac{6\alpha}{n^3} + \frac{2\beta}{n^2} - \frac{\gamma}{n}+ \delta \approx 0.685.
\end{eqnarray}
Then, once the value on $n$ is fixed, we obtain a constraint on the values of $\alpha$, $\beta$, $\gamma$ and $\delta$.\\
For example, if we choose $n=2$, we obtain the following set of values:
 $\alpha = 1$,  $\beta = 2$, $\gamma = 0.5$ and $\delta = 0.685$. Another possible set is: $\alpha = 0.5$,  $\beta = 1$, $\gamma = 0.2$ and $\delta = 0.66$.\\
 If we consider the case with $n=3$, we obtain the following set of values: $\alpha = 1$,  $\beta = 2$, $\gamma = 0.5$ e $\delta = 0.63$.\\
 If we consider $n=4$, we obtain the following set of values $(\alpha,\beta,\gamma,\delta)=(1,1,1,0.90375)$.

 In Table \ref{tab:q1_values}, we have the values of $q_{1,0}$ for the different values of $n$ we considered and for the different values of the parameters we derived. Moreover, we considered $\Omega_{m0}=0.315$ and $\Omega_{k0}=0.01$. 
\begin{table}[h!]
\centering
\begin{tabular}{|c|c|c|}
\hline
$n$ & $(\alpha,\beta,\gamma,\delta)$ & $q_{1,0}$ \\
\hline
2 & (1, 2, 0.5, 0.685) & -0.185 \\
2 &  (0.5, 1, 0.2, 0.66) & -0.185 \\
3 & (1, 2, 0.5, 0.63) & -0.299 \\
4 & (1, 1, 1, 0.90375) & -0.356 \\
\hline
\end{tabular}
\caption{Values of $q_{1,0}$ for different values of $n$ and corresponding sets of parameters $(\alpha,\beta,\gamma,\delta)$, with $\Omega_{m0}=0.315$ and $\Omega_{k0}=0.01$.}
\label{tab:q1_values}
\end{table}
For all the cases considered, we obtained a negative value of $q$, indicating an accelerated expansion of the Universe. 

\subsection{Interacting Case}
We now consider the presence of interaction between Dark Sectors.\\
We now extend our analysis by considering the possibility of an interaction between the dark sectors. This idea refers to scenarios where dark matter (DM) and dark energy (DE) are not entirely independent but may exchange energy or momentum. Such a coupling is often motivated by attempts to address the so-called coincidence problem, namely why the energy densities of DM and DE are of the same order today despite their different evolutionary histories. Allowing for an interaction modifies the standard cosmological dynamics and can leave distinctive observational imprints, such as changes in the expansion history, deviations in structure formation, or shifts in the cosmic microwave background (CMB) anisotropies. These models have therefore been widely studied as possible alternatives or extensions to the concordance $\Lambda$CDM framework.

In the presence of such a coupling, the conservation equations for DE and DM are modified as follows:
\begin{eqnarray}
\dot{\rho}_{D} + 3H \rho_{D} (1+\omega_{D}) &=& -Q, \label{eq:DE_cons}\\
\dot{\rho}_{m} + 3H \rho_{m} &=& Q, \label{eq:DM_cons}
\end{eqnarray}
where $Q$ specifies the rate of energy transfer between the two sectors. In general, $Q$ may be a function of several cosmological quantities, including the Hubble parameter $H$, the deceleration parameter $q$, and the energy densities $\rho_{m}$ and $\rho_{D}$, i.e. $Q=Q(\rho_{m},\rho_{D},H,q)$. A variety of choices for this function have been considered in the literature. In our study, we adopt the phenomenological form
\begin{equation}
Q = 3 d^{2} H \rho_{m}, \label{Q}
\end{equation}
where $d^{2}$ is a dimensionless constant quantifying the strength of the interaction, often called the transfer rate or coupling parameter~\cite{ref144d,ref145d,ref146d}.

Observational analyses combining different cosmological probes — such as the Gold SNe~Ia sample, CMB data from WMAP, and BAO measurements from SDSS — suggest that $d^{2}$ should be positive defined and it should assume a small value. This outcome is in agreement with the requirements imposed by the cosmic coincidence problem as well as with thermodynamical considerations~\cite{ref147d}. Additional constraints from CMB anisotropy studies and galaxy cluster observations further indicate the range $0 < d^{2} < 0.025$~\cite{ref148d}. More generally, the parameter is usually considered within $[0,1]$, with the special case $d^{2}=0$ reducing to the standard non-interacting FRW cosmology. It is worth stressing that many other functional forms of $Q$ have been proposed in the literature, each leading to different phenomenological consequences.

The expression of the energy density of DE is the same as in the non-interacting case.\\ 
For the energy density of DM, from the continuty equation for the interacting case, we obtain the following expression of $\rho_m$ as function of the scale factor:
\begin{eqnarray}
    \rho_{m,I}= \rho_{m0}a^{-3(1-d^2)}.
\end{eqnarray}
Using the expression of $a(t)$ we have chosen in this paper, we can write:
\begin{eqnarray}
    \rho_{m,I}(t)&=& \rho_{m0}b_0^{-3(1-d^2)}t^{-3n(1-d^2)}\nonumber \\
    &=&\frac{\rho_{m0}}{b_0^{3(1-d^2)}t^{3n(1-d^2)}},
\end{eqnarray}
from which we obtain:
\begin{eqnarray}
    \rho_{m,I}(t)&=& 3H_0^2\Omega_{m0}b_0^{-3(1-d^2)}t^{-3n(1-d^2)}\nonumber \\
    &=&\frac{3H_0^2\Omega_{m0}}{b_0^{3(1-d^2)}t^{3n(1-d^2)}}.
\end{eqnarray}

The expression of $\rho_{m,I}$ for the interacting case as function of the redshift is given by
\begin{eqnarray}
    \rho_{m,I}(z)&=& \rho_{m0}(1+z)^{3(1-d^2)},
\end{eqnarray}
which leads to
\begin{eqnarray}
    \rho_{m,I}(z)&=& 3H_0^2\Omega_{m0}(1+z)^{3(1-d^2)}.
\end{eqnarray}

\begin{figure}[htbp]
    \centering
    \includegraphics[width=0.7\textwidth]{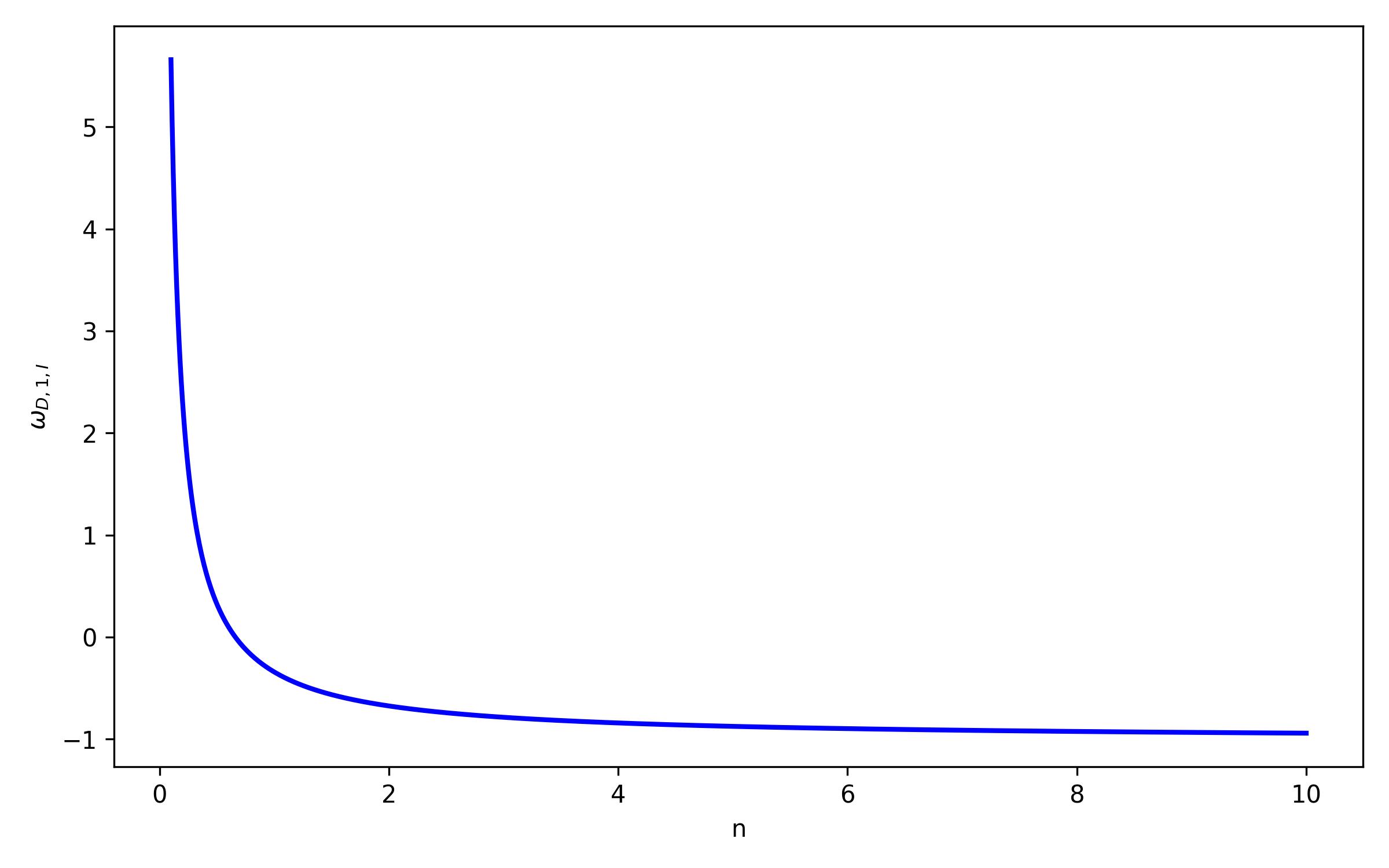} 
    \caption{Plot of the expression of $\omega_{D,1,I}$ for $n \in \left[0.1-10\right] $.}
    \label{2}
\end{figure}

In this case, the Hubble parameter squared is given by:
\begin{eqnarray}
    H_{1,I}^2(z)&=& H_0^2\left\{  \Omega_{m0}(1+z)^{3(1-d^2)}+\Omega_{k0}(1+z)^2+ \frac{1}{H_0^2} \left( -\frac{6\alpha}{n} + 2\beta - \gamma n + \delta n^2 \right)b_0^{2\over n}\,(1+z)^{2\over n}\right\}\nonumber\\
    &=& H_0^2\left\{  \Omega_{m0}(1+z)^{3(1-d^2)}+\Omega_{k0}(1+z)^2+ \left(-\frac{6\alpha}{n^3} + \frac{2\beta}{n^2} - \frac{\gamma}{n}+ \delta \right)\,(1+z)^{2\over n}\right\}. \label{unical3}
\end{eqnarray}
From the continuity equation given in Eq. (\ref{eq:DE_cons}), we obtain the following expression for the EoS parameter for DE:
\begin{eqnarray}
    \omega_{D,1,I} = -1 +(1+z) \frac{\rho'_{D,1}}{3\rho_{D,1}}- \frac{Q}{3H\rho_{D,1}},
\end{eqnarray}
where in this case a prime indicates a derivative with respect to $z$.\\
The term $\frac{Q}{3H\rho_{D,1,I}}$ is given by:
\begin{eqnarray}
    \frac{Q}{3H\rho_{D,1}} &=& d^2\left(\frac{\rho_{m,I}}{\rho_{D,1}}\right)\nonumber \\
    &=&\frac{d^2\Omega_{m0}(1+z)^{3(1-d^2)}}{\left(-\frac{6\alpha}{n^3} + \frac{2\beta}{n^2} - \frac{\gamma}{n}+ \delta \right)(1+z)^{2\over n}},
\end{eqnarray}
where we used the fact that $\rho_{m0} = 3H_0^2\Omega_{m0}$.\\
The final expression of the EoS parameter is then given by:
\begin{eqnarray}
    \omega_{D,1,I}(z) = -1 + \frac{2}{3n}- \frac{d^2\Omega_{m0}(1+z)^{3(1-d^2)}}{\left(-\frac{6\alpha}{n^3} + \frac{2\beta}{n^2} - \frac{\gamma}{n}+ \delta \right)(1+z)^{2\over n}}\label{ruth}.
\end{eqnarray}
In the limiting case of $d^2=0$, we obtain the expression derived for the non-interacting case.\\
In Fig. (\ref{2}) we plot the behavior of the EoS parameter of DE $\omega_D$ obtained in Eq. (\ref{ruth})  in the range of values of $n \in \left[0.1-10\right] $. We considered $d^2=0.02$. Moreover, we have chosen $\Omega_{m0}=0.315$ and we have that $-\frac{6\alpha}{n^3} + \frac{2\beta}{n^2} - \frac{\gamma}{n}+ \delta $ is always 0.685 for all the combinations of values we have considered. 
The present day value of the EoS is given by:
\begin{eqnarray}
    \omega_{D,1,I,0} = -1 + \frac{2}{3n}- \frac{d^2\Omega_{m0}}{-\frac{6\alpha}{n^3} + \frac{2\beta}{n^2} - \frac{\gamma}{n}+ \delta }.
\end{eqnarray}
Considering the values we are taking into account, we have that 
\begin{eqnarray}
    \frac{d^2\Omega_{m0}}{-\frac{6\alpha}{n^3} + \frac{2\beta}{n^2} - \frac{\gamma}{n}+ \delta }\approx 0.00920.
\end{eqnarray}
Therefore, the presence of interaction does not affect the final value of the EoS considerably.\\
We now want to calculate the expression of the pressure of DE for this case. \\
From the continuity equation for DE for the non-interacting case, we obtain:
\begin{eqnarray}
    p_{D,1,I}= -\rho_{D,1}  +(1+z) \frac{\rho_{D,1}'}{3H}-\frac{Q}{3H}.
\end{eqnarray}
The term $\frac{Q}{3H}$ is given by:
\begin{eqnarray}
    \frac{Q}{3H} = d^2\rho_{m,I} = 3d^2H_0^2\Omega_{m0}(1+z)^{3(1-d^2)}.
\end{eqnarray}
Therefore, we obtain:
\begin{eqnarray}
p_{D,1,I}(z)  =
\left(  \frac{2}{n}-3  \right) \left(-\frac{6\alpha}{n^3} + \frac{2\beta}{n^2} - \frac{\gamma}{n}+ \delta \right)H_0^2(1+z)^{2\over n}-3d^2H_0^2\Omega_{m0}(1+z)^{3(1-d^2)}.
\end{eqnarray}
In the limiting case of $d^2=0$, we recover the same result of the non-interacting case.\\
We now want to obtain the final expression of the deceleration parameter $q_{1,I}$.\\
We still use the general expression:
\begin{eqnarray}
    q_{1,I}(z)= -1+\left(\frac{1+z}{2}\right)\frac{1}{h_{1,I}^2(z)}\frac{dh_{1,I}^2(z)}{dz}.
\end{eqnarray}
Using the expression of the Hubble parameter squared obtained in Eq. (\ref{unical3}), we can write:
\begin{eqnarray}
    \frac{dh_{1,I}^2(z)}{dz} &=& 3(1-d^2)  \Omega_{m0}(1+z)^{3(1-d^2)-1} +2\Omega_{k0}(1+z)\nonumber \\
    &&+ \frac{2}{nH_0^2} \left( -\frac{6\alpha}{n} + 2\beta - \gamma n + \delta n^2 \right)b_0^{2\over n}\,(1+z)^{\frac{2}{n}-1}\nonumber \\
    &=& 3(1-d^2)\Omega_{m0}(1+z)^{3(1-d^2)}+2\Omega_{k0}(1+z)^2\nonumber \\
    &&+\frac{2}{n}\left(-\frac{6\alpha}{n^3} + \frac{2\beta}{n^2} - \frac{\gamma}{n}+ \delta \right)\,(1+z)^{2\over n}.
\end{eqnarray}
Therefore, the final expression of $q_{1,I}$ is given by:
\begin{eqnarray}
    q_{1,I}(z)
    &=&-1+\frac{1}{2}\cdot \left[  3(1-d^2)\Omega_{m0}(1+z)^{3(1-d^2)}+2\Omega_{k0}(1+z)^2\right.\nonumber \\
    &&\left.\,\,\,\,\,\,\,\,\,\,\,\,\,\,\,\,\,\,\,\,\,\,\,\,\,+ \frac{2}{n}\left(-\frac{6\alpha}{n^3} + \frac{2\beta}{n^2} - \frac{\gamma}{n}+ \delta \right)\,(1+z)^{\frac{2}{n}}  \right]\times \nonumber \\
    &&\left[  \Omega_{m0}(1+z)^{3(1-d^2)}+\Omega_{k0}(1+z)^2+  \left(-\frac{6\alpha}{n^3} + \frac{2\beta}{n^2} - \frac{\gamma}{n}+ \delta \right)\,(1+z)^{\frac{2}{n}}  \right]^{-1}.
\end{eqnarray}

The present day value of $q_{1,I}$ can be written as: 
\begin{eqnarray}
    q_{1,I,0}
    &=&-1+\frac{1}{2}\cdot \left[  3(1-d^2)\Omega_{m0}+2\Omega_{k0}+ \frac{2}{n}\left(-\frac{6\alpha}{n^3} + \frac{2\beta}{n^2} - \frac{\gamma}{n}+ \delta \right)\, \right]\times \nonumber \\
    &&\left[  \Omega_{m0}+\Omega_{k0}+  \left(-\frac{6\alpha}{n^3} + \frac{2\beta}{n^2} - \frac{\gamma}{n}+ \delta \right)\,  \right]^{-1}.
\end{eqnarray}

 In Table \ref{tab:q1I_all_cases_compact}, we have the values of $q_{1,I,0}$ for the different values of $n$ we considered and for the different values of the parameters we derived. Moreover, we considered $\Omega_{m0}=0.315$ and $\Omega_{k0}=0.01$. 
\begin{table}[h!]
\centering
\begin{tabular}{|c|c|c|}
\hline
$n$ & $(\alpha,\beta,\gamma,\delta)$ & $q_{1,I,0}$ \\
\hline
2 & (1, 2, 0.5, 0.685)  & -0.194 \\
2 &  (0.5, 1, 0.2, 0.66) & -0.194 \\
3 & (1, 2, 0.5, 0.63) & -0.309 \\
4 & (1, 1, 1, 0.90375) & -0.365 \\
\hline
\end{tabular}
\caption{Values of $q_{1,I,0}$ for different values of $n$ and corresponding sets of parameters $(\alpha,\beta,\gamma,\delta)$, with $\Omega_{m0}=0.315$, $\Omega_{k0}=0.01$ and $d^2=0.02$.}
\label{tab:q1I_all_cases_compact}
\end{table}
For all the cases considered, we obtained a negative value of $q$, indicating an accelerated expansion of the Universe.

\section{Second Model}
We now consider the second model of this paper.\\
Using the scale factor defined in Eq. (\ref{scale}), we obtain the following expression for the energy density of DE:
\begin{eqnarray}
\rho_{D,2}(z)
&=&3\left[-\frac{6\alpha}{n^3} + \frac{2(\zeta+\beta)}{n^2} - \frac{\gamma}{n}+ \delta \right] H_0^{2}(1+z)^{2\over n}.
\end{eqnarray}
In this case, the Hubble parameter squared can be written as:
\begin{eqnarray}
    H_2^2(z)= H_0^2\left\{  \Omega_{m0}(1+z)^3+\Omega_{k0}(1+z)^2+  \left[-\frac{6\alpha}{n^3} + \frac{2(\zeta+\beta)}{n^2} - \frac{\gamma}{n}+ \delta \right]\,(1+z)^{2\over n}\right\}.
\end{eqnarray}
The expression of the EoS parameter of DE $\omega_D$ is the same of the first model:
\begin{eqnarray}
    \omega_{D,2} = -1+\frac{2}{3n}.
\end{eqnarray}
Following the same procedure of the first model, we obtain that the pressure of DE is given by:
\begin{eqnarray}
p_{D,2}(z) &=& \left(\frac{2}{3n}-1\right) \rho_{D,2}(z) \nonumber\\
&=& \left( \frac{2}{n}-3\right) 
 \left[-\frac{6\alpha}{n^3} + \frac{2(\zeta+\beta)}{n^2} - \frac{\gamma}{n}+ \delta \right] H_0^{2}(1+z)^{2\over n}.
\end{eqnarray}

The final expression of the deceleration parameter $q_2(z)$ is given by:
\begin{eqnarray}
    q_2(z)&=& -1 + \frac{1}{2}\cdot \left\{ 3\Omega_{m0}(1+z)^3+2\Omega_{k0}(1+z)^2\right.\nonumber \\
    &&\left.\,\,\,\,\,\,\,\,\,\,\,\,\,\,\,\,\,\,\,\,\,\,\,\,\,\,\,+ \frac{2}{n}\left[-\frac{6\alpha}{n^3} + \frac{2(\zeta+\beta)}{n^2} - \frac{\gamma}{n}+ \delta \right] \,(1+z)^{\frac{2}{n}}  \right\}\times \nonumber \\
    &&\,\,\,\,\,\,\,\,\,\,\,\,\,\,\,\,\,\,\,\,\,\,\,\,\,\,\,\,\left\{   \Omega_{m0}(1+z)^3+\Omega_{k0}(1+z)^2\right.\nonumber \\
    &&\left.\,\,\,\,\,\,\,\,\,\,\,\,\,\,\,\,\,\,\,\,\,\,\,\,\,\,\,\,\,+ \left[-\frac{6\alpha}{n^3} + \frac{2(\zeta+\beta)}{n^2} - \frac{\gamma}{n}+ \delta \right] \,(1+z)^{\frac{2}{n}}  \right\}^{-1}.
\end{eqnarray}
Therefore, we have that the present-day value of $q_2$ is given by:
\begin{eqnarray}
    q_{2,0}&=& -1 + \frac{1}{2}\cdot \left\{ 3\Omega_{m0}+2\Omega_{k0}+ \frac{2}{n}\left[-\frac{6\alpha}{n^3} + \frac{2(\zeta+\beta)}{n^2} - \frac{\gamma}{n}+ \delta \right]  \right\}\times \nonumber \\
    &&\,\,\,\,\,\,\,\,\,\,\,\,\,\,\,\,\,\,\,\,\,\,\,\,\,\,\,\,\left\{   \Omega_{m0}+\Omega_{k0}+ \left[-\frac{6\alpha}{n^3} + \frac{2(\zeta+\beta)}{n^2} - \frac{\gamma}{n}+ \delta \right]   \right\}^{-1}.
\end{eqnarray}
In Table \ref{tab2}, we write the values of $q_{2,0}$ for the values of $n$ we studied and for different combinations of values of the parameters we considered. We have also considered $\Omega_{m0} =0.685$ and $\Omega_{k0} =0.001$. For all the cases considered, we obtained a negative value of $q$, indicating an accelerated expansion of the Universe.

\begin{table}[h!]
\centering
\begin{tabular}{|c|c|c|}
\hline
$n$ & $(\alpha,\beta,\zeta,\gamma,\delta)$ & $q_{2,0}$ \\
\hline
2 & \multicolumn{1}{c|}{(2, 0.5, 1, 0.2, 1.535)} & -0.185 \\
3 & \multicolumn{1}{c|}{(1, 1, 1, 2, 1.13)} & -0.299 \\
4 & \multicolumn{1}{c|}{(1, 0.13, 0.1, 1, 1)} & -0.356 \\
\hline
\end{tabular}
\caption{Values of $q_{2,0}$ for different values of $n$, with $\Omega_{m0}=0.315$, $\Omega_{k0}=0.01$, and corresponding sets of parameters $(\alpha,\beta,\zeta,\gamma,\delta)$. Moreover, we consider $\Omega_{m0} =0.685$ and $\Omega_{mk0} =0.001$.}
\label{tab2}
\end{table}

We now want to find the evolutionary form of the fractional energy density of DE. \\
Using the expression of $\rho_{D,2}$, we can write $\Omega_{D,2}$ as follow:
\begin{eqnarray}
\Omega_{D,2} &=& \frac{\rho_{D,2}}{3H^2}\nonumber\\
&=&\left[-\frac{6\alpha}{n^3} + \frac{2(\zeta+\beta)}{n^2} - \frac{\gamma}{n}+ \delta \right].
\end{eqnarray}
i.e. $\Omega_{D,2}$ is a constant depending on the values of the parameters of the energy density and on the power law index of the scale factor we have chosen. \\
Therefore, the evolutionary form of $\Omega_{D,2}$ is $\Omega_{D,2}'=0$, i.e. we have that the fraction of DE relative to the total critical density does not change over time.\\
We have then that the present day value of $\Omega_{D,2}$ is given by:
\begin{eqnarray}
\Omega_{D,2,0} 
&=&\left[-\frac{6\alpha}{n^3} + \frac{2(\zeta+\beta)}{n^2} - \frac{\gamma}{n}+ \delta \right].
\end{eqnarray}
We can obtain some hints about the values of the parameters of the model using the value of $\Omega_{D0}(z)$. \\
We now want that, at present time,  $\Omega_{D,2,0}\approx 0.685$, therefore we should have:
\begin{eqnarray}
-\frac{6\alpha}{n^3} + \frac{2(\zeta+\beta)}{n^2} - \frac{\gamma}{n}+ \delta  \approx 0.685.
\end{eqnarray}
Therefore, once the value of $n$ is fixed, we obtain a constraint on the values of $\alpha$, $\beta$, $\gamma$, $\delta$ and $\zeta$.\\
For example, if we choose $n=2$, we obtain the following set of possible values: $\alpha = 2$,  $\zeta = 1$,  $\beta = 0.5$,  $\gamma = 0.2$ and  $\delta = 1.535$.\\
If we choose $n=3$, we obtain this possible set of values:
$\alpha=1$, $\beta=1$, $\zeta=1$, $\gamma=2$, $\delta=1.13$.\\
If we choose $n=4$, we obtain that $\alpha=1$, $\beta=0.13$, $\gamma=1$, $\delta=1$ e $\zeta =0.1$. \\

\subsection{Interacting Case}
We now consider the interacting case.\\
The expression of $\rho_D$ is the same obtained in the non-interacting case.\\
The Hubble parameter squared for this case can be written as:
\begin{eqnarray}
    H_{2,I}^2(z)&=& H_0^2\left\{  \Omega_{m0}(1+z)^{3(1-d^2)}+\Omega_{k0}(1+z)^2\right.\nonumber \\
    &&\left.\,\,\,\,\,\,\,\,\,\,\,\,+ \left[-\frac{6\alpha}{n^3} + \frac{2(\zeta+\beta)}{n^2} - \frac{\gamma}{n}+ \delta \right] \,(1+z)^{2\over n}\right\}.
\end{eqnarray}

From the continuity equation given in Eq. (\ref{eq:DE_cons}), we obtain the following expression for the EoS parameter for DE:
\begin{eqnarray}
    \omega_{D,2,I} = -1 +(1+z)\frac{\rho'_{D,2}}{3\rho_{D,2}}- \frac{Q}{3H\rho_{D,2}}.
\end{eqnarray}
The term $\frac{Q}{3H\rho_{D,2}}$ is given by:
\begin{eqnarray}
    \frac{Q}{3H\rho_{D,2}} &=& d^2\left(\frac{\rho_{m,I}}{\rho_{D,2}}\right)\nonumber \\
    &=&\frac{d^2\Omega_{m0}(1+z)^{3(1-d^2)}}{\left[-\frac{6\alpha}{n^3} + \frac{2(\zeta+\beta)}{n^2} - \frac{\gamma}{n}+ \delta \right](1+z)^{2\over n}},
\end{eqnarray}
where we used the fact that $\rho_{m0} = 3H_0^2\Omega_{m0}$.\\
The final expression of the EoS parameter is then given by:
\begin{eqnarray}
    \omega_{D,2,I} = -1 +\frac{2}{3n}- \frac{d^2\Omega_{m0}(1+z)^{3(1-d^2)}}{\left[-\frac{6\alpha}{n^3} + \frac{2(\zeta+\beta)}{n^2} - \frac{\gamma}{n}+ \delta \right](1+z)^{2\over n}}.
\end{eqnarray}
In the limiting case of $d^2=0$, we obtain the expression derived for the non-interacting case.\\
The behavior of $ \omega_{D,2,I}$ is the same as $\omega_{D,1,I}$ since the values of the parameters involved in the interacting term assume similar values.\\
The present-day value of the EoS is given by:
\begin{eqnarray}
    \omega_{D,2,I,0} = -1 + \frac{2}{3n}- \frac{d^2\Omega_{m0}}{-\frac{6\alpha}{n^3} + \frac{2(\zeta+\beta)}{n^2} - \frac{\gamma}{n}+ \delta }.
\end{eqnarray}
As in the other model, the interaction term does not affect the value of the EoS considerably.

We now want to calculate the expression of the pressure of DE for this case. \\
From the continuity equation for DE for the non-interacting case, we obtain:
\begin{eqnarray}
    p_{D,2,I}= -\rho_{D,2} +(1+z) \frac{\rho_{D,2}'}{3H}-\frac{Q}{3H}.
\end{eqnarray}
The term $\frac{Q}{3H}$ is given by:
\begin{eqnarray}
    \frac{Q}{3H} = d^2\rho_{m,I} = 3d^2H_0^2\Omega_{m0}(1+z)^{3(1-d^2)}.
\end{eqnarray}

Therefore, we obtain:
\begin{eqnarray}
p_{D,2,I}(z)  =
 \left(  \frac{2}{n}-3  \right)\left[-\frac{6\alpha}{n^3} + \frac{2(\zeta+\beta)}{n^2} - \frac{\gamma}{n}+ \delta \right]H_0^2(1+z)^{2\over n}-3d^2H_0^2\Omega_{m0}(1+z)^{3(1-d^2)}.
\end{eqnarray}
In the limiting case of $d^2=0$, we recover the same result of the non-interacting case.\\
For the deceleration parameter $q$, we use the general definition introduced before:
\begin{eqnarray}
    q_{2,I} = -1+\left(\frac{1+z}{2}\right)\frac{1}{h_{2,I}^2(z)}\frac{dh_{2,I}^2(z)}{dz}.
\end{eqnarray}
Using the expression of $H_{2,I}^2(z)$, we obtain:
\begin{eqnarray}
    \frac{dh_{2,I}^2(z)}{dz} &=& 3(1-d^2)  \Omega_{m0}(1+z)^{3(1-d^2)-1} +2\Omega_{k0}(1+z)\nonumber \\
    &&+ \frac{2}{n} \left[-\frac{6\alpha}{n^3} + \frac{2(\zeta+\beta)}{n^2} - \frac{\gamma}{n}+ \delta \right](1+z)^{\frac{2}{n}-1},
\end{eqnarray}
where $h_{2,I}^2= H_{2,I}^2/H_0^2$.\\
Therefore, the final expression of $q_{2,I}$ is given by:
\begin{eqnarray}
    q_{2,I}&=& -1 + \frac{1}{2}\cdot \left\{ 3(1-d^2)\Omega_{m0}(1+z)^{3(1-d^2)}+2\Omega_{k0}(1+z)^2\right.\nonumber \\
    &&\left.\,\,\,\,\,\,\,\,\,\,\,\,\,\,\,\,\,\,\,\,\,\,\,\,\,\,\,\,\,+ \frac{2}{n}\left[-\frac{6\alpha}{n^3} + \frac{2(\zeta+\beta)}{n^2} - \frac{\gamma}{n}+ \delta \right](1+z)^{\frac{2}{n}}  \right\}\times \nonumber \\
    &&\,\,\,\,\,\,\,\,\,\,\,\,\,\,\,\,\,\,\,\,\,\,\,\,\,\,\left\{   \Omega_{m0}(1+z)^{3(1-d^2)}+\Omega_{k0}(1+z)^2\right.\nonumber \\
    &&\left.\,\,\,\,\,\,\,\,\,\,\,\,\,\,\,\,\,\,\,\,\,\,\,\,\,\,\,\,\,\,\,+ \left[-\frac{6\alpha}{n^3} + \frac{2(\zeta+\beta)}{n^2} - \frac{\gamma}{n}+ \delta \right]\,(1+z)^{\frac{2}{n}}  \right\}^{-1}.
\end{eqnarray}
The present-day value of $q_{2,I}$ is given by:
\begin{eqnarray}
    q_{2,I,0}&=& -1 + \frac{1}{2}\cdot \left\{ 3(1-d^2)\Omega_{m0}+2\Omega_{k0}\right.\nonumber \\
    &&\left.\,\,\,\,\,\,\,\,\,\,\,\,\,\,\,\,\,\,\,\,\,\,\,\,\,\,\,\,\,+ \frac{2}{n}\left[-\frac{6\alpha}{n^3} + \frac{2(\zeta+\beta)}{n^2} - \frac{\gamma}{n}+ \delta \right] \right\}\times \nonumber \\
    &&\,\,\,\,\,\,\,\,\,\,\,\,\,\,\,\,\,\,\,\,\,\,\,\,\,\,\left\{   \Omega_{m0}+\Omega_{k0}\right.\nonumber \\
    &&\left.\,\,\,\,\,\,\,\,\,\,\,\,\,\,\,\,\,\,\,\,\,\,\,\,\,\,\,\,\,\,\,+ \left[-\frac{6\alpha}{n^3} + \frac{2(\zeta+\beta)}{n^2} - \frac{\gamma}{n}+ \delta \right]  \right\}^{-1}.
\end{eqnarray}
In Table \ref{tab2I}, we write the values of $q_{2,I,0}$ for the different values of $n$ we choose and the combinations of the values of the parameters we considered. Moreover, we considered $\Omega_{m0} =0.685$, $\Omega_{k0} =0.001$  and $d^2=0.02$. For all the cases considered, we obtained a negative value of $q$, indicating an accelerated expansion of the Universe. 

\begin{table}[h!]
\centering
\begin{tabular}{|c|c|c|}
\hline
$n$ & $(\alpha,\beta,\zeta,\gamma,\delta)$ & $q_{2,I,0}$ \\
\hline
2 & \multicolumn{1}{c|}{(2, 0.5, 1, 0.2, 1.535)} & -0.194 \\
3 & \multicolumn{1}{c|}{(1, 1, 1, 2, 1.13)} & -0.309 \\
4 & \multicolumn{1}{c|}{(1, 0.13, 0.1, 1, 1)} & -0.365 \\
\hline
\end{tabular}
\caption{Values of $q_{2,I,0}$ for different values of $n$, with $\Omega_{m0}=0.315$, $\Omega_{k0}=0.01$, $d^2=0.02$, and corresponding sets of parameters $(\alpha,\beta,\zeta,\gamma,\delta)$. }
\label{tab2I}
\end{table}
For the present day value of the deceleration parameter of the second model, in both non-interacting and interacting cases, we find the same values we found for the first model as it was expected, since the values of the parameters we are considering lead to equal values of the quantities involved.

\section{Age of the Present Universe}
The determination of the Universe’s age is one of the cornerstone results of modern cosmology. Within the standard $\Lambda$CDM paradigm, this age can be estimated by integrating the Friedmann equations backward in time from the present epoch to the initial singularity, using the observed Hubble expansion rate. The calculated value is highly sensitive to the cosmological parameters, particularly the current Hubble constant $H_0$, the matter density parameter $\Omega_m$, and the dark energy contribution $\Omega_D$. Current measurements, for instance those from the Planck mission, suggest an age of $t_0 \simeq 13.8 \, \mathrm{Gyr}$, with uncertainties at the percent level. Nevertheless, the ongoing $H_0$ tension between early- and late-time observations implies that slightly different values of $H_0$ could shift the inferred cosmic age by several hundred million years. Moreover, alternative cosmological frameworks—such as interacting dark energy models, modifications of general relativity, or scenarios with non-zero spatial curvature—can also affect the theoretical estimate of $t_0$, providing an additional observational handle to discriminate between competing models of the Universe.\\
The age of universe can be determined thanks to:
\begin{eqnarray}
    t_0 -t = -\int_{t_0}^t dt = \int_0^z\frac{dz'}{(1+z')H(z')} .\label{ageuni}
\end{eqnarray}
We now calculate the present age of the Universe for the models we studied in this paper.\\
Eq. (\ref{ageuni}) does not have an analytical solution for these models, therefore we solve it numerically considering a value of $z=100$.\\
For the first model, in the case with $n=2$ and $\alpha = 1$,  $\beta = 2$, $\gamma = 0.5$ and $\delta = 0.685$ and for the case with $n=2$ and $\alpha = 0.5$,  $\beta = 1$, $\gamma = 0.2$ and $\delta = 0.66$, we obtain $t_0-t\approx 12.951$ Gyr for the non interacting case while $t_0-t\approx 13.191 $ Gyr for the interacting case.\\
Instead, for $n=3$ and $\alpha = 1$,  $\beta = 2$, $\gamma = 0.5$ e $\delta = 0.63$, we obtain $t_0-t \approx 13.266$ Gyr for the non-interacting case, while we obtain $t_0-t \approx 13.522$ Gyr for the interacting case.\\
For $n=4$ and $(\alpha,\beta,\gamma,\delta)=(1,1,1,0.90375)$, we obtain $t_0-t \approx 13.410$ Gyr for the non-interacting case and $t_0-t \approx 13.672$ Gyr for the interacting case.\\
 We now consider the second DE energy density model we studied.\\
 For $n=2$ and $\alpha = 2$,  $\zeta = 1$,  $\beta = 0.5$,  $\gamma = 0.2$ and  $\delta = 1.535$, we obtain $t_0-t \approx 12.951$ Gyr for the non-interacting case and $t_0-t \approx 13.191$ Gyr for the interacting case.\\
When $n=3$ and $\alpha=1$, $\beta=1$, $\zeta=1$, $\gamma=2$, $\delta=1.13$, we obtain $t_0-t \approx 13.266$ Gyr for the non-interacting case and $t_0-t \approx 13.522$ Gyr for the interacting case.\\
For $n=4$ and $\alpha=1$, $\beta=0.13$, $\gamma=1$, $\delta=1$ e $\zeta =0.1$, we obtain $t_0-t \approx 13.410$ Gyr for the non interacting case and $t_0-t \approx 13.672$ Gyr for the interacting case. \\
We observe we find the same values for both models, as it was expected, since the values of the parameters we used lead to the same values for the quantities involved in the calculations.

\section{Conclusions}
In this paper, we studied two dark energy (DE) models involving higher-order derivatives of the Hubble parameter $H$. These models can be considered as generalizations of previously studied DE scenarios.

By assuming a power-law form for the scale factor $a(t)$, we derived expressions for the DE energy density, DE pressure, equation-of-state (EoS) parameter, deceleration parameter, and the evolutionary behavior of the fractional DE energy density for both non-interacting and interacting dark sector scenarios. For the interacting case, we adopted a coupling term proportional to the dark matter (DM) energy density.

We determined several combinations of parameter values for the models corresponding to power-law indices $n=2$, $n=3$, and $n=4$. Using these parameter sets, we calculated the present-day age of the Universe. The resulting values slightly differ from the observationally inferred age of $\approx 13.8$ Gyr, even if they are quite close. Nevertheless, we observe that the calculated age increases with $n$, and that interacting scenarios consistently yield higher values compared to the non-interacting cases.

Future works can be devoted to studying the search for best fit values for the involved parameters. Furthermore, these models can be studied considering other scale factors in order to better understand their behavior.


\begin{thebibliography}{99}
\bibitem{cmb2} D.N. Spergel, et al., Astrophys. J. Suppl. Ser. \textbf{148}, 175 (2003).
\bibitem{cmb1} C. L. Bennett, et al., Astrophys. J. \textbf{148}, 1 (2003).
\bibitem{sn4} P. de Bernardis et al, Nature \textbf{404}, 955 (2000).
\bibitem{sn2} S. Perlmutter et al., Astrophys. J. \textbf{517}, 565 (1999).
\bibitem{sds1} M. Tegmark et al., Phys. Rev. D \textbf{69}, 103501 (2004).

\bibitem{sds4} J.K. Adelman-McCarthy, et al., Astrophys. J. Suppl. Ser. \textbf{175}, 297 (2008).
\bibitem{sds3} K. Abazajian et al., Astron. J. \textbf{129}, 1755 (2005).
\bibitem{planck} Planck Collaboration, P.A.R. Ade et al.\ 2013, arXiv:1303.5076
\bibitem{xray} S.W. Allen et al., Mon. Not. Roy. Astron. Soc. \textbf{353}, 457 (2004).
\bibitem{cosm4} P.J.E. Peebles, B. Ratra, Reviews of Modern Physics, \textbf{75}, 559 (2003).
\bibitem{cosm3} V. Sahni, A. Starobinsky, Int. J. Mod. Phy. D \textbf{9}, 373 (2000).
\bibitem{cosm5} T. Padmanabhan, Phys. Rep. \textbf{380}, 235 (2003).
\bibitem{rev2} S. D. Odintsov, D. S-C. Gomez, G. S. Sharov, Eur. Phys. C  \textbf{77}, 862 (2017).
\bibitem{twothirds} H.V. Peiris, et al., Astrophys. J. Suppl. Ser. \textbf{148}, 213 (2003).
\bibitem{quint1} B. Ratra, B., Peebles, P.J.E., Phys. Rev. D, \textbf{37}, 3406 (1988).
\bibitem{quint4} P.J.E. Peebles, B. Ratra, Astrophys. J. Lett., \textbf{325}, L17 (1988).
\bibitem{quint5} C. Wetterich, Nucl. Phys. B \textbf{302}, 668 (1988).
%\bibitem{quint6} R.R. Caldwell, R. Dave, P.J. Steinhardt, Phys. Rev. Lett. \textbf{80}, 1582 (1998).

\bibitem{kess5} A. Pasqua, A. Khodam-Mohammadi, M. Jamil, R. Myrzakulov, Astrophys. Space Sci. \textbf{340}, 199 (2012).
\bibitem{kess4} C.A. Picon, T. Damour, V. Mukhanov, Phys. Lett. B \textbf{458}, 209 (1999).
\bibitem{kess3} T. Chiba, T. Okabe, M. Yamaguchi, Phys. Rev. D \textbf{62}, 023511 (2000).


%\bibitem{addition1} K. Bamba, S. Capozziello, S. Nojiri, S. D. Odintsov, Astrophys Space Sci  \textbf{342}, 155 (2012).
\bibitem{tac2} T. Padmanabhan, Phys. Rev. D \textbf{66}, 021301 (2002).
\bibitem{tac3} T. Padmanabhan, T.R. Choudhury, Phys. Rev. D \textbf{66}, 081301 (2002).
\bibitem{tac4} A. Pasqua, M. Jamil, R. Myrzakulov B.  Majeed,  Physica Scripta \textbf{86}, 045004 (2012).
\bibitem{pha2} S. Nojiri, S.D. Odintsov, Phys. Lett. B \textbf{565}, 1-9 (2003).
\bibitem{pha5} L.P. Chimento, R. Lazkoz, Phys. Rev. Lett. \textbf{91}, 211301 (2003).
\bibitem{pha6} B. Boisseau, G. Esposito-Farese, D. Polarski, Alexei A. Starobinsky,Phys. Rev. Lett. \textbf{85}, 2236 (2000).
\bibitem{dil2} M. Gasperini, F. Piazza, G. Veneziano, Phys. Rev. D \textbf{65}, 023508 (2002).
\bibitem{dil1} N. Arkani-Hamed, P. Creminelli, S. Mukohyama, M. Zaldarriaga, J. Cosmol. Astropart. Phys. \textbf{4}, 1 (2004).
\bibitem{qui3} S. Nojiri, S.D. Odintsov, S. Tsujikawa, Phys. Rev. D \textbf{71}, 063004 (2005).
\bibitem{qui8} W. Zhao, Y. Zhang, Phys. Rev. D \textbf{73}, 123509 (2006).
\bibitem{qui10} H. Mohseni Sadjadi, M. Alimohammadi, Phys. Rev. D \textbf{74}, 043506 (2006).
\bibitem{qui12} M.R. Setare, E.N. Saridakis, J. Cosmol. Astropart. Phys. \textbf{0809}, 026 (2008).
\bibitem{cgas1} A. Kamenshchik, U. Moschella, V. Pasquier, Physics Letters B \textbf{511}, 265 (2001).
\bibitem{cgas2} M.C. Bento, O. Bertolami, A.A. Sen, Phys. Rev. D \textbf{66}, 043507 (2002).
\bibitem{cgas3} M.R. Setare, European Physical Journal C \textbf{52}, 689 (2007).
\bibitem{ade2} R.G. Cai, Phys. Lett. B \textbf{657}, 228 (2007).
\bibitem{ade1} H. Wei, R. G. Cai, Physics Letters B \textbf{660}, 113 (2008).


\bibitem{holo2} L. Susskind, J. Math. Phys. \textbf{36}, 6377 (1995).
\bibitem{holo1} W. Fischler, L. Susskind, (arXiv:hep-th/9806039).
\bibitem{holo5} Q.G. Huang, M. Li, J. Cosmol. Astropart. Phys. \textbf{8}, 13 (2004).
\bibitem{li} M. Li, Phys. Lett. B \textbf{603}, 1 (2004).




\bibitem{coh1} A. Cohen, D. Kaplan, A. Nelson, Phys. Rev. Lett. \textbf{82}, 4971 (1999).
\bibitem{gaoprimo} C. Gao, F. Wu, X. Chen, Y. G. Shen, Phys. Rev. D \textbf{79}, 043511 (2009).
\bibitem{go1} L.N. Granda,  International Journal of Modern Physics D \textbf{18}, 1749 (2009).
\bibitem{go4} L.N. Granda, A. Oliveros, Physics Letters B \textbf{671}, 199 (2009).
\bibitem{go5} L.N. Granda, A. Oliveros, Physics Letters B \textbf{669}, 275 (2008).
\bibitem{cons2} X. Zhang, F.Q. Wu, Phys. Rev. D \textbf{76}, 023502 (2007).
\bibitem{cons1} K. Enqvist, S. Hannestad, M. S. Sloth, J. Cosmol. Astropart. Phys. \textbf{2}, 4 (2005).
\bibitem{cons4} X. Zhang, Phys. Rev. D \textbf{79}, 103509 (2009).
\bibitem{cons3} X. Zhang, F.Q. Wu, Phys. Rev. D \textbf{72}, 043524 (2005).
\bibitem{cons5} Y. Wang, L. Xu, Phys. Rev. D \textbf{81}, 083523 (2010).
\bibitem{cons7} H.C. Kao, W.L. Lee, F.L. Lin, Phys. Rev. D \textbf{71}, 123518 (2005).
\bibitem{cons6} S.M.R. Micheletti, J. Cosmol. Astropart. Phys. \textbf{4}, 9 (2010).
\bibitem{cons9} J. Shen, B. Wang, E. Abdalla, R. K. Su, Physics Letters B \textbf{609}, 200 (2005).
\bibitem{cons8} B. Feng, X. Wang,  X. Zhang, Physics Letters B \textbf{607}, 35 (2005).

\bibitem{hde2} M.R. Setare, Physics Letters B \textbf{642}, 421 (2006).
\bibitem{hde1} S. Nojiri, S. D. Odintsov, E. N. Saridakis, R. Myrzakulov, Nuclear Physics B \textbf{950} 114850 (2020).
\bibitem{hde7} M.R. Setare, European Physical Journal C \textbf{50}, 991 (2007).
\bibitem{hde10} S. Nojiri, S. D. Odintsov, Eur. Phys. J. C \textbf{77} 528 (2017).
\bibitem{hde12} B. Guberina, R. Horvat, H. Stefancic, J. Cosmol. Astropart. Phys. \textbf{5}, 1 (2005).
\bibitem{hde13} Y. Gong, Phys. Rev. D \textbf{70}, 064029 (2004).
\bibitem{hde17} M. Jamil, E.N. Saridakis, M.R. Setare, Physics Letters B \textbf{679}, 172 (2009).
\bibitem{hde18} M. Jamil, M.U. Farooq, M. A. Rashid, European Physical Journal C \textbf{61}, 471 (2009).
\bibitem{hde19} A. Sheykhi, Astrophys. J. \textbf{27}, 025007 (2010).
\bibitem{hde22} H.M. Sadjadi, M. Jamil, General Relativity and Gravitation \textbf{43}, 1759 (2011).
\bibitem{hde23} M.R. Setare, M. Jamil, J. Cosmol. Astropart. Phys. \textbf{2}, 10 (2010).
\bibitem{hde30} M.R. Setare, M. Jamil, Physics Letters B \textbf{690}, 1 (2010).
\bibitem{hde26} A. Sheykhi, Physics Letters B \textbf{681}, 205 (2009).
\bibitem{hde24} M.R. Setare, M. Jamil, General Relativity and Gravitation \textbf{43}, 293 (2011).
\bibitem{hde28} M.R. Setare, S. Shafei, J. Cosmol. Astropart. Phys. \textbf{9}, 11 (2006).
\bibitem{hde33} E. Elizalde, S. Nojiri, S.D. Odintsov, P. Wang, Phys. Rev. D \textbf{71}, 103504 (2005).
\bibitem{hde32} K. Karami, M.S. Khaledian, F. Felegary, Z. Zarmi, Physics Letters B \textbf{686}, 216 (2010).
\bibitem{hde35} X. Zhang, Phys. Rev. D \textbf{74}, 103505 (2006).
\bibitem{hde34} M.U. Farooq, M. Jamil, M. A. Rashid, Int. J. Theor. Phys. \textbf{49}, 2334 (2010).
\bibitem{saridakis22} J. Lu, E. N. Saridakis, M. R. Setare, L. Xu,  J. Cosmol. Astropart. Phys. \textbf{03}, 031 (2010).
\bibitem{saridakis11}  M.R. Setare, E.N. Saridakis, Phys. Lett. B \textbf{671}, 331  (2009).
\bibitem{YYS} Z. Yao, G. Ye, A. Silvestri 	arXiv:2508.01378 
https://doi.org/10.48550/arXiv.2508.01378





\bibitem{modelhigher} S. Chen, J. Jing, Physics Letters B \textbf{679}, 144 (2009).
\bibitem{gohnde} L.N. Granda, A. Oliveros, Phys. Lett. B \textbf{669}, 275 (2008)
\bibitem{altri2} S. Nojiri, S.D. Odintsov, Phys. Rev. D \textbf{72}, 023003 (2005).
\bibitem{altri1} D. Jain, A. Dev, Phys. Lett. B \textbf{633}, 436 (2006).

\bibitem{altri3} S. Capozziello, V. Cardone, E. Elizalde, S. Nojiri, S.D. Odintsov, Phys. Rev. D \textbf{73}, 043512 (2006).


\bibitem{ref146d} W. Zimdahl, D. Pavon, Gen. Rel. Grav. \textbf{35}, 413 (2003).
\bibitem{ref145d} A. Sheykhi, M. Jamil, Phys. Lett. B \textbf{694}, 284 (2011).
\bibitem{ref144d} L. Amendola, D. Tocchini-Valentini, Phys. Rev. D \textbf{64}, 043509 (2001).

\bibitem{ref147d} C. Feng, et al., Phys. Lett. B \textbf{665}, 111 (2008).
\bibitem{ref148d} K. Ichiki, et al., J. Cosmol. Astropart. Phys. \textbf{06}, 005 (2008).


\end{thebibliography}
\end{document}